\documentclass[twocolumn, tighten]{aastex7}

\usepackage[T1]{fontenc}
\usepackage{microtype}
\usepackage{newtxtext}
\usepackage[varvw]{newtxmath}

\usepackage{amsmath} 
\usepackage{subcaption}
\usepackage{comment}

\usepackage{natbib}

\begin{document}

\title{Tailored mass estimators for Milky Way dwarf Spheroidals}

\author[]{Sofia L. Splawska}\email{splawska@cmu.edu}
\affiliation{McWilliams Center for Cosmology and Astrophysics, Department of Physics, Carnegie Mellon University, Pittsburgh, PA 15213, USA}

\author[]{Rapha\"el Errani}\email{errani@cmu.edu}
\affiliation{McWilliams Center for Cosmology and Astrophysics, Department of Physics, Carnegie Mellon University, Pittsburgh, PA 15213, USA}

\author[]{Jorge Pe\~narrubia}\email{jorpega@roe.ac.uk}
\affiliation{Institute for Astronomy, University of Edinburgh, Royal Observatory, Blackford Hill, Edinburgh EH9 3HJ, UK}

\author[]{Matthew G. Walker}\email{mgwalker@cmu.edu}
\affiliation{McWilliams Center for Cosmology and Astrophysics, Department of Physics, Carnegie Mellon University, Pittsburgh, PA 15213, USA}

\begin{abstract}

Assuming spherical symmetry and dynamical equilibrium within a given gravitational potential, a dwarf spheroidal (dSph) galaxy’s globally averaged stellar velocity dispersion depends entirely on the shape of its stellar density profile.  Thus, the dynamical inference of a dSph’s gravitational potential is necessarily sensitive to assumptions about that shape.  Relaxing standard assumptions, we fit flexible stellar density models to observations of the Milky Way’s known dSph satellites.  Considering various choices for the density profile shape and spatial extent of a host dark matter halo, we use the virial theorem to propagate observational uncertainties about the shapes of the inferred dSph stellar density profiles to uncertainties in the inferred dynamical masses.  We find that the observed structural diversity of the Milky Way dSph population implies a large range of potential systematic errors (up to factors of 10) associated with standard dynamical mass estimators.  We show that accounting for these observational and systematic uncertainties can significantly alter the appearance and behavior of dSph dynamical scaling relations, including enclosed dynamical mass vs. stellar mass and the Radial Acceleration Relation.

\end{abstract}


\section{Introduction}

Probing the extremes of small size, low luminosity, and high dynamical density, dwarf galaxies are essential probes of cosmology and galaxy formation physics. However, the motion of stars within these systems cannot be explained by Newtonian gravity alone if all the mass sourcing the gravitational potential is contributed by the observed, luminous mass. In the cold dark matter picture of cosmology, this discrepancy is explained by dark matter, dwarf galaxies being the most dark matter-dominated objects. 

Robust mass constraints for dwarf Spheroidal (dSph) galaxies are critical to exploiting their discriminatory power with regards to the nature of dark matter.  While the observed luminosity can be straightforwardly converted to baryonic mass, the gravitational potential---including any contribution from dark matter---must be inferred dynamically from the distribution of stellar positions and velocities. In most cases, observations resolve only positions on the sky and velocities along the line of sight.  Given this incomplete information about the stellar phase space, it is common to use simple dynamical mass estimators of the form $M({<}r_0)=\mu r_0\langle\sigma_{\rm los}^2\rangle/G$, where $M({<}r_0)$ is the inferred dynamical mass enclosed within a sphere of radius $r_0$ centered on the galaxy, and $\langle\sigma_{\rm los}^2\rangle$ is the squared global velocity dispersion. The latter quantity has the benefit of being robust for systems with few stars and, crucially, is independent of the velocity anisotropy.

While the coefficient $\mu$ generally depends on the shapes and sizes of the density profiles for both stellar and dark matter components, it is common to adopt fixed values to construct `simple' mass estimators. For example, \cite{walker2009universal} use $\mu=2.5$ for $r_0=R_{\rm h}$, the projected half-light radius; \cite{wolf2010accurate} use $\mu=3.0$ for $r_0=r_{\mathrm{h}}$, the 3D half-light radius, which corresponds to $\mu=4.0$ for $r_0=R_{\rm h}$. After examining systematics associated with these estimators, \cite{errani2018systematics} suggest $\mu=3.5$ for $r_0=1.8R_{\rm h}$ as a value that minimizes systematic errors when the stellar component follows a \cite{plummer11} profile, marginalizing over different dark matter profiles.  

A wealth of new photometric survey data of dwarf spheroidals enables careful study of the shapes of their stellar density profiles \citep[e.g.,][]{munoz18}.  In a companion paper, Walker et al. (2026, in preparation), we fit flexible models to the stellar density profiles of the known Milky Way satellites, relaxing standard assumptions about those profiles in order to infer the shapes observationally.  The result is a range of shapes, characterized by a diversity of inner and outer density profile slopes, motivating the use of mass estimators that can accommodate this diversity.  

In this work, we examine the range of dynamical coefficients $\mu$ associated with the range of shapes inferred for the known Milky Way dSphs under different assumptions of the dSph's internal gravitational potential. Extending the formalism of  \cite{errani2018systematics}, we use the projected virial theorem \citep{agnello2012virial} to develop dynamical mass estimates that are `tailored' to the individual stellar density profile observed for each galaxy.

Given the critical role that mass estimators play in studying dSph scaling relations (see, e.g., \citealt{2012AJ....144....4M, 2019ARA&A..57..375S}), we then examine how the adoption of our tailored masses affects the behaviors of such scaling relations, using the enclosed dynamical mass - stellar mass relation and the Radial Acceleration Relation of \cite{mcgaugh2016radial}, \cite{lelli2017one} as examples. 

The paper is organized as follows: in Section \ref{sec methods} we review the projected virial theorem and the dependence of $\mu$ on the shapes of the stellar and dark matter density profiles. In Section \ref{sec data}, we describe the data and fitting procedure used and motivate the use of a flexible stellar density profile. In Section \ref{sec results} we study the systematics of the mass estimator for mass follows light models and stellar tracers embedded in dark matter halos. In Section \ref{sec scaling relations} we show how these tailored mass estimates affect scaling relations and discuss; we conclude in Section \ref{sec conclusion}.

\section{Dynamical masses from the virial theorem}
\label{sec methods}

Following \cite{errani2018systematics}, we adpot a mass estimate based on the projected spherical virial theorem that relates the 1D projected component of the kinetic and potential energy tensors \citep{agnello2012virial}:

\begin{equation}
\label{virial}
    2K_\mathrm{\mathrm{los}} = - W_\mathrm{\mathrm{los}}.
\end{equation}
The left-hand side of the above equation is identical to the  surface-density weighted line-of-sight velocity dispersion $\langle \sigma_{\mathrm{los}}^2 \rangle$,
\begin{equation}
2K_{\mathrm{los}} = 2\pi \int_0^{\infty} \Sigma_{\star}(R) \, \sigma_{\mathrm{los}}^2(R)\,  R \, \mathrm{d}R \equiv \langle \sigma_{\mathrm{los}}^2 \rangle,
\label{eq:2klos}
\end{equation}
where $\Sigma_{\star}(R)$ is the projected stellar number density at projected radius $\mathrm R$, normalized so that $2\pi \int_0^{\infty} R \,\Sigma_{\star}(R)\, \mathrm{d}R = 1$.
The right-hand side of Equation \ref{virial} can be computed from the 3D stellar number density $\nu_{\star}(r)$ and the enclosed dynamical mass profile $M({<}r)$:
\begin{equation}
W_{\mathrm{los}} = -\frac{4\pi G}{3} \int_0^{\infty} r \, \nu_{\star}(r) \, M({<}r) \, \mathrm{d}r ~,
\label{eq:wlos}
\end{equation}
where $\nu_{\star}(r)$ is normalized so that $4\pi\int_0^{\infty} r^2 
\,\nu_{\star}(r) 
\,\mathrm d r = 1$.
We can use Equations \ref{virial}-\ref{eq:wlos} to calculate the potential energy term $W_{\mathrm{los}}$ from the assumed $M({<}r)$ and $\nu_{\star}(r)$ alone, linking the observed line-of-sight kinematics $\langle \sigma_{\mathrm{los}} \rangle = - W_{\mathrm{los}}$  with the underlying potential. Note that here, unlike in the Jeans equations, the velocity dispersion anisotropy plays no role, allowing us to bypass the mass-anisotropy degeneracy. 
This allows us to calculate the dynamical mass enclosed within a 3D radius $r_0$:
\begin{equation}
\label{mass 1}
    M({<}r_0) = \frac{\mu r_0 \langle \sigma^2_{\mathrm{tot}} \rangle}{\mathrm G},
\end{equation}
where $\mu$ is a dimensionless coefficient.
We compute $\mu$ through Equations \ref{eq:wlos} and \ref{mass 1} as 
\begin{equation}
\label{mu 1}
    \mu \equiv -\frac{GM({<}r_0)}{r_0 W_{\mathrm{los}}},
\end{equation}
which, for a given $\nu_{\star}(r)$ and $M({<}r)$, is a function of the radius $r_0$.
Note that the value of $\mu(r_0)$ depends only on the \emph{shapes} of the stellar number density $\nu_{\star}(r)$ and the underlying distribution of dynamical mass $M({<}r)$, but not on their absolute amplitudes.

It is common to adopt simple mass estimators that are based on Equation \ref{mass 1} but which assign a constant value to $\mu$.  For example, the \cite{walker2009universal} and \cite{wolf2010accurate} estimators both assume the line-of-sight velocity dispersion profile is approximately constant with projected radius, enabling (via the Jeans equations) constraints on $M({<}R_{\rm h})$ without explicitly specifying a shape for $M({<}r)$.  Assuming the stellar density profile takes the \citet{plummer11} form, \citet{walker2009universal} obtain a fixed value $\mu=2.5$ for $r_0=R_{\rm h}$. Assuming a constant velocity anisotropy and allowing for a variety of standard stellar density profiles, \citet{wolf2010accurate} obtain $\mu\approx3$ for $r_0=r_{-3}$, the radius where the log-slope decreases to a value of $d\log\nu/d\log r=-3$, and show that this result is equivalent to $\mu\approx 4$ for $r_0=R_{\rm h}$. It is important to note that while most of the luminous `classical' dSphs are confirmed to have approximately flat line-of-sight velocity dispersion profiles \citep{walker2007velocity}, the shapes of velocity dispersion profiles for the less-luminous `ultrafaint' dSphs are weakly constrained and in most cases completely unknown, giving little justification for applying the estimators with fixed $\mu$.

Equation \ref{mass 1} holds for any equilibrium combination of $M({<}r)$ and $\nu_\star(r)$ and does not make any demands on the shape of the velocity dispersion profile or anisotropy. Following \cite{2012MNRAS.419..184A} we denote the 3D radius $r_0$ within which we aim to estimate the enclosed dynamical mass as a multiple of the 2D half-light radius, $r_0 = \lambda R_{\mathrm h}$.  With this notation, the mass is as follows:
\begin{equation}
\label{mass 2}
    M({<}\lambda R_{\mathrm h}) = \frac{\mu \lambda R_{\mathrm h} \langle \sigma^2_{\mathrm{tot}} \rangle}{\mathrm G}.
\end{equation}

In this work, we use the above equation to refer to the mass enclosed within the projected 2D stellar half-light radius by setting $\lambda = 1$.

\section{Observed diversity of stellar density profiles}
\label{sec data}

\begin{figure*}[t] % 'h' means "here"
    \centering
    \includegraphics[width=1.0\textwidth]{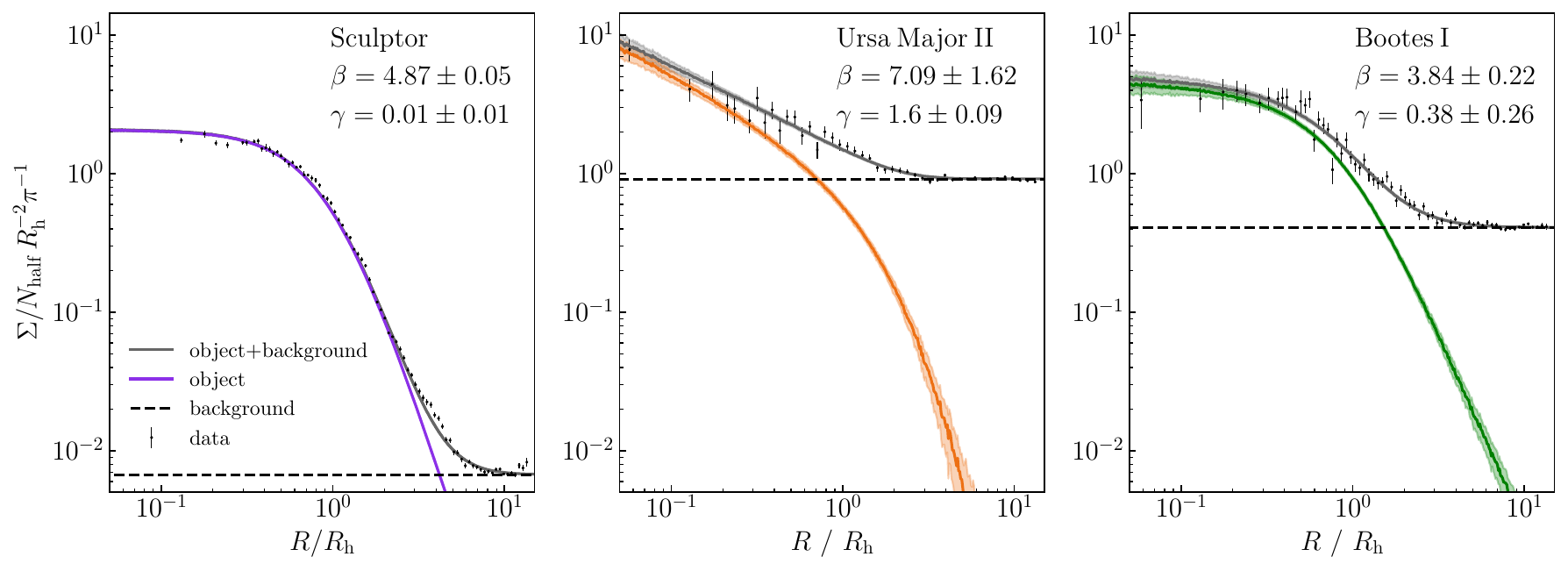} % Adjust width as needed
    \caption{The stellar density profiles of Milky Way satellites display a
remarkable diversity in shapes, which can be accurately fitted by the
flexible $2\beta\gamma$ 3D stellar density distribution of
Equation \ref{eq:2bg}.  Shown are binned stellar surface number density profiles $\Sigma$ as a function of
the 2D radius $R$ for the three Milky Way satellites Sculptor, Ursa Major II,
and Bootes I, normalized by the average number of stars within the half-light radius. Black points with error bars shown binned projected stellar number density profiles (see Section \ref{sec data} for details). A black horizontal line
indicates a the central value of the gradient-fitted foreground and colored bands indicate
the 1-sigma uncertainty region of projected $2\beta\gamma$
profiles. }
    \label{SB fig}
\end{figure*}

The Local Volume Database (LVD) of \citet{Pace2024arXiv241107424P}\footnote{\url{https://github.com/apace7/local_volume_database}}  contains up-to-date published measurements of the quantities---half-light radius and line-of-sight velocity dispersion---that serve as inputs to dynamical mass estimators.  However, the collection of published half-light radii has limited utility for our purposes.  First, they are obtained using a wide variety of instrumentation and analysis methods, with heterogeneity introducing potential systematic errors when studying the dwarf galaxy population as a whole.  Second, most published half-light radii are estimated assuming the stellar surface density follows one of a few simple profiles of fixed shape --- usually exponential or Plummer models.  Systematic errors associated with such restrictive modeling assumptions are rarely quantified (but see \citealt{cicuendez18,munoz18,wang19}) and, in our perception, not widely appreciated.  In any case, the mass in Equation \ref{mass 2} depends on the assumed shape of the stellar density profile, and so is sensitive to any assumptions thereabout.

In an effort to mitigate these problems, we perform a homogeneous analysis of all known Milky Way dwarf galaxies, fitting a flexible stellar density model to data drawn from public catalogs of major sky surveys.  We refer the reader to Walker et al. (2026, in preparation), henceforth Paper II, for full details of this procedure and complete results for all individual galaxies.  Briefly, we select all point sources projected within square fields of side length 10 half-light radii (previously-published, as listed in the LVD) of all known dwarf Milky Way companions observed in any of the Sloan Digital Sky Survey (SDSS, Data Release 9), Pan-STARRS (PS1, Data Release 1), Dark Energy Survey (DES, Data Release 2), Dark Energy Camera Legacy Survey (DECaLS, Data Release 9), DECALS Local Volume Exploration Survey (DELVE, Data Release 2), and/or \textit{Gaia} (Early Data Release 3).  We use model isochrones to define color/magnitude criteria for selecting candidate members of each dwarf galaxy.  

Independently for each survey catalog, we infer parameters of the stellar density profile, $\nu(r)$, by modeling its 2D projection 
\begin{equation}
    \Sigma(R)=2\displaystyle\int_{R}^{\infty}\frac{r\,\nu(r)\,dr}{\sqrt{r^2-R^2}}
    \label{eq:bigsigma}
\end{equation}
against the distribution of projected positions of candidate members. Here, $\Sigma(R)= N_\star\Sigma_\star(R)$ and $\nu(r)=N_\star\nu_\star(r)$, where $N_\star$ is the total number of stars observed and the star subscript denotes the normalized profiles (see Section \ref{sec methods}). Assuming that the number of stars observed within the area element $d^2\vec{R}$ at projected position $\vec{R}$ is drawn from a Poisson distribution with expectation value $\Sigma(\vec{R})d^2\vec{R}$, then given model $M$, the observed data $D$ have a probability that satisfies
\begin{equation}
    \ln P(D | M)= \displaystyle\sum_{i=1}^{N_{\rm{pix}}}\ln \Sigma(\vec{R_i})-\displaystyle\int_{\rm field}\Sigma(\vec{R})\,d^2\vec{R}
    \label{eq:lnp}
\end{equation}
where the integral is taken over the observed field.  In practice, we approximate the integral in Equation \ref{eq:lnp} as $\int_{\rm field}\Sigma(\vec{R})\,d^2\vec{R}\approx A_{\rm pix}\sum_{i=1}^{N_{\rm pix}}\Sigma(\vec{R}_i)$, where we discretize the field of area $400 R_{\rm h}^2$ (where $R_{\rm h}$ is the published 2D halflight radius from the LVD) using a grid of $250\times 250$ pixels.  This choice implies that individual pixels have area $A_{\rm pix}\sim (0.08 R_{\rm h})^2$ and that $\sim 500$ pixels resolve the region within 1 (published) $R_{\rm h}$.  We mask pixels containing known structures (other objects cataloged in the LVD) and artifacts from bright stars (we also  exclude all stars within masked pixels from the sum in the first term on the right-hand side of Equation \ref{eq:lnp}).  

In order to accommodate a wide range of shapes for the stellar density profile, we adopt the double power-law model 
\begin{equation}
\nu(r) = \nu_0 \left( \frac{r}{a} \right)^{-\gamma} \left[ 1 + \left( \frac{r}{a} \right)^{2} \right]^{(\gamma - \beta)/2},\label{eq:2bg}
\end{equation}
which has slope $d\log\nu/d\log r=-\gamma$ at $r\ll a$ and $d\log\nu/d\log r=-\beta$ at $r\gg a$, where $a$ is the scale radius.  This model is a generalization of the standard Plummer profile, for which $(\beta,\gamma)=(5,0)$, and a special case of the $\alpha\beta\gamma$ profile \citep{hernquist1990analytical,zhao1996analytical} with $\alpha=2$.  We use `$2\beta\gamma$' to denote the model expressed by Equation \ref{eq:2bg}. We will refer to stellar profiles with constant central density, $\gamma \approx 0$ as `cored' and $\gamma>0$ as `cuspy'. While Paper II allows for elliptically-flattened morphology, given the spherical symmetry inherent in the virial mass estimator (Eqs. \ref{eq:2klos} and \ref{eq:wlos}), we assume $\nu(r)$ is spherically symmetric and, via Equation \ref{eq:bigsigma}, its projection $\Sigma(R)$ has circular symmetry. Therefore, we adopt the set of results from Paper II that assume circular symmetry, with ellipticity set to zero. Fitting spherical as opposed to elliptical models does not introduce an overall bias in the resulting $\beta$, $\gamma$, or $R_{\rm h}$. Specific cases where the flattened and spherical models result in meaningfully different structural parameters will be discussed individually.

In addition to $\beta$, $\gamma$, scale radius $a$, and scale density $\nu_0$, free parameters of our model include the target galaxy's central coordinates and the fraction of observed stars that are members of the target galaxy as opposed to a foreground component that is allowed to have a density gradient.  With the likelihood function specified by Equation \ref{eq:lnp}, we use the nested sampling algorithm \citep{skilling04} MultiNest \citep{feroz09} to infer the posterior probability distribution for model parameters (see Paper II for complete details).  We adopt uniform priors on $\beta$ and $\gamma$ over the ranges $[3,10]$ and $[0,2]$, respectively, a uniform prior on $\log_{10} [a/\textrm{arcmin}]$ between $[-1,2.5]$ and a uniform prior for $\log_{10}[\Sigma_0/\mathrm{arcmin^{-2}}]$ between $[-10,10]$. The results include posterior probability distributions for each galaxy's total stellar mass, which is computed assuming a \citet{kroupa02} stellar mass function in order to account for stars fainter than the adopted magnitude limit, with stellar mass related to filter magnitude according to the MIST isochrone \citep{dotter16} MIST isochrone \citep{dotter16} with old age (between 10-14 Gyr) and the galaxy's LVD-tabulated mean metallicity (see Paper II for complete details).

For each survey catalog and fit to each galaxy, our companion paper provides a rough signal-to-noise ratio that quantifies significance of the local overdensity due to galaxy members, and a $\chi^2$ statistic that quantifies goodness of the $2\beta\gamma$ model fit to a binned stellar surface density profile (even though the actual fits are to the distribution of discrete stellar positions and not to the binned profiles).  We discard results for which either $S/N<3$ or $\chi^2/N_{\rm bin}>3$.  For galaxies with results for multiple survey catalogs passing these filters, we adopt the results from the case with largest $S/N$.

Across the Milky Way dwarf galaxy population, our fits imply a diverse range of structural parameters $\beta$ and $\gamma$ as defined in Equation \ref{eq:2bg}.  To illustrate this, Figure \ref{SB fig} shows the projected stellar number density of the Sculptor, Ursa Major II, and Bootes I satellite galaxies as well as their $2\beta\gamma$ fits. They range from cored, Plummer-like profiles (Sculptor, $\beta = 4.47 \pm0.05, \: \gamma = 0.01\pm0.01$) to significantly cuspy ones with steep outer slopes (Ursa Major II, $\beta = 7.09\pm1.62, \: \gamma= 1.60\pm0.09$), to very shallow outer slopes (Bootes I, $\beta = 3.84\pm0.22, \: \gamma = 0.38\pm0.26$) and span a wide range of luminosities (${\sim} 10^3-10^7 \,L_{\odot}$). The three galaxies are representative of a striking diversity in profile shapes observed throughout our sample of Milky Way dwarf spheroidals. We will refer to these systems throughout as examples of how the unique stellar density features influence estimates of their enclosed dynamical mass.

\section{Results}
\label{sec results}

Given the observationally constrained stellar density profiles $\nu_{\star}(r)$, in order to calculate the dynamical coefficient $\mu$, we must specify the shape of the enclosed mass profiles, $M({<}r)$, which source the gravitational potential (Equation \ref{mu 1}).  We consider two scenarios: one in which the shape of the stellar tracer density profile $\nu_{\star}(r)$ is identical to the shape of the density profile that gives rise to the enclosed dynamical mass $M({<}r)$ (`mass follows light'), and one in which the stellar component is embedded within a dark matter halo that dominates the dynamical mass profile and whose density profile need not resemble that of the stars.

\subsection{Mass follows light models}
\label{MFL}
In this section, we apply the formalism of Section \ref{sec methods} to calculate values for the dynamical coefficient $\mu$ under the assumption that mass follows light (see, e.g., \citealt{1976ApJ...204...73I, 1986AJ.....92...72R}). For mass follows light models, the total dynamical mass is given by
\begin{equation}
M({<}r) =  L \Upsilon_\mathrm{dyn} \int_0^r 4\pi s^2\nu_\star(s) ds,
\label{MFL mass}
\end{equation}
where $L$ and $\Upsilon_\mathrm{dyn}$ denote the total luminosity and the dynamical mass-to-light ratio, respectively. Some authors have used full dynamical models to conclude that the Carina, Fornax, Sculptor, and Sextans dSphs are consistent with mass follows light models (\cite{diakogiannis2017novel, kowalczyk2019schwarzschild, lokas2009mass}, see also \cite{kleyna2002dark}).

 \begin{figure}[h] % 'h' means "here"
    \centering
    \includegraphics[width=1.0\columnwidth]{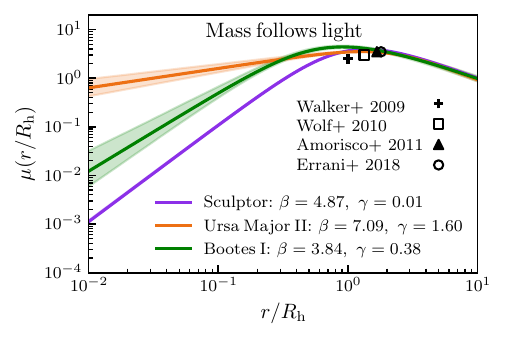} % Adjust width as needed
    \caption{For mass follows light models, the dynamical coefficient $\mu$ (see Equation \ref{mu 1}) depends on the radius $r$ used when defining the enclosed mass $M({<}r)$. Shown is the variation in $\mu$ vs. enclosing radius between the three galaxies shown in Figure \ref{SB fig}. $\mu$ is plotted for the best fit parameters (solid lines) with 1-sigma shading from uncertainties on $R_{\mathrm{h}}$, distance, $\beta$, and $\gamma$. Black symbols show simple mass estimators taken from the literature.}
    \label{MFL fig}
\end{figure}

  Figure \ref{MFL fig} shows the resulting relationship between $\mu$ and the 3D radius $r$ equal to the 2D radius within which the enclosed mass is defined for the three galaxies introduced in Figure \ref{SB fig}. For small radii ($r \ll R_{\mathrm{h}}$) the value of $\mu$ is very sensitive to the shape of the stellar density profile $\nu_\star(r)$. For example, for Sculptor, which is well-fitted by a cored stellar density profile $(\gamma = 0.01 \pm 0.01)$ and a Plummer-like outer slope $(\beta = 4.87 \pm0.05)$, we find that $\mu \approx 0.11 $ at $r/R_{\mathrm{h}} = 0.1$. For Ursa Major II, well-approximated by a cuspy $2\beta\gamma$ profile with $\beta = 7.09\pm1.62$ and $\gamma = 1.60\pm0.09$, $\mu \approx 1.59 $ at the same radius. In contrast, for $r \sim R_{\mathrm{h}}$, we find that the inferred enclosed masses (see Equation \ref{mass 2}) are relatively insensitive to the assumptions made about the stellar density profile, recovering the the `core fitting' result of \citet{1986AJ.....92...72R}. We note that the functional dependence of $\mu$ on $R$ is sensitive to the value of $\gamma$ and less so on the value of $\beta$.

We also plot values of $\mu$ from various published simple mass estimators. We note that these generally \textit{do not} assume mass follows light. We include the \cite{walker2009universal} estimator ($\lambda = 1,\,\mu = 2.5$), the \cite{wolf2010accurate} estimator ($\lambda = 4/3,\,\mu = 3$), the minimum variance estimator of \cite{errani2018systematics}
 ($\lambda = 1.8,\,\mu = 3.5$), and the \cite{amorisco2011phase} estimator based on distribution function models of Michie-King profiles  in NFW halos ($\lambda = 1.7,\,\mu = 5.8$). These estimators fall in the region of minimum variance between the three $2 \beta \gamma$ profiles. Therefore, for mass follows light models, many simple mass estimators are accurate regardless of the details of the stellar density profile.

\subsection{Stellar tracers embedded in dark matter halos}
\label{stars + dm}
Next, we calculate the coefficient $\mu$ for massless stellar tracers embedded in dark matter halos, where the profile of the mass sourcing the potential is not directly observable. We defer to \ref{sec:not massless} the case where stars contribute to the potential. Two additional unknowns arise here compared to the mass follows light case treated in Section \ref{MFL}: (1) the shape of the underlying dark matter halo, and (2) the (radial) size of the dark matter halo with respect to the stellar component. As before, we model the stellar tracers using the $2\beta\gamma$ profile to tailor the mass to the galaxy's individual stellar density profile. For the dark matter density distributions, we consider a cuspy \cite{1990ApJ...356..359H} profile 

\begin{equation}
    \rho_{\mathrm{cusp}}(r) = \frac{M_{\mathrm{DM}}}{2\pi} \frac{a}{r (r + a)^{3}},
    \label{cusp density}
\end{equation}

and a cored \cite{dehnen1993family} profile 

\begin{equation}
    \rho_{\mathrm{core}}(r) = \frac{3 M_{\mathrm{DM}}}{4\pi}  \frac{a}{ (r + a)^4}.
    \label{core density}
\end{equation}
In the above equations, $a$ denotes the scale radius and $M_{\mathrm{DM}}$ is the total halo mass. These examples are motivated by the inner shapes of dark matter halos in cold dark matter-only cosmological simulations (cuspy halos, see, e.g., \citealt{navarro1997universal}), as well as in simulations of cold dark matter with baryons or self-interacting dark matter (cored halos, see, e.g., \citealt{onorbe2015forged}, \citealt{{vogelsberger2012subhaloes}} respectively). \cite{errani2018systematics} show that virial theorem mass estimates (see Equation \ref{mass 2}) vary significantly between these two scenarios for a Plummer tracer profile. Here we re-examine this variation in the context of the observed diversity of tracer density profiles.

We parameterize the relative size of the dark matter halo with respect to the stellar component using
the ratio $R_\mathrm{h}/ r_\textnormal{max}$, where $r_\textnormal{max}$ denotes the radius of
maximum circular velocity of the underlying dark matter halo $v_{\text{max}} = \left( G M_{\mathrm{DM}}({<}r_{\text{max}})/r_{\text{max}}\right)^{1/2}$, easily accessible in simulations. $R_\mathrm{h}/ r_\textnormal{max}$ represents how deeply
the stellar component is segregated within the dark matter halo. $R_{\mathrm h}/r_{\mathrm{max}} \ll 1 $ denotes stellar systems that are very deeply embedded in the surrounding dark matter halo, and $R_{\mathrm h}/r_{\mathrm{max}} \approx 1 $ those in which the characteristic radii of the stellar and dark matter profiles coincide (for example, in tidally limited dwarf galaxies, see \citealt{errani2022structure}). For (cuspy) Hernquist profiles, $r_\textnormal{max}$ is related to the scale radius $a$ through $r_\textnormal{max} =a$, while for (cored) Dehnen profiles,
$r_\textnormal{max} = 2a$. For a given $\nu_\star(r)$ and $M_{\mathrm{DM}}({<}r)$, the dimensionless factor $\mu$ is a function of the segregation $R_\mathrm{h}/r_\mathrm{max}$. However, $R_\mathrm{h}/r_\mathrm{max}$ is generally unknown. 
Hydrogen cooling constraints suggest that dwarf galaxies form in dark matter halos with masses larger than ${\sim}10^8$ (see, e.g., \citealt{2020MNRAS.498.4887B}), which corresponds to a spatial scale $r_{\mathrm{max}} \sim 10^3 \:\mathrm{pc}$ \citep{2016MNRAS.460.1214L}, while $R_{\mathrm h}$ is observed to vary from $10-10^3 \: \mathrm{pc}$ in Milky Way dSphs \citep{Pace2024arXiv241107424P}. Therefore, we consider $0.01 \leq R_\mathrm{h}/r_\mathrm{max} \leq1$, as a plausible range (see also \citealt{2022MNRAS.510.3967R, 2023MNRAS.520.1630K, 2025arXiv250203679S} for the relation between halo size and galaxy size in cosmological simulations). Evolutionary effects can alter the segregation after formation of the system; this and other caveats will be discussed in Section \ref{sec discuss}.

\begin{figure*}[t] % 'h' means "here"
    \centering
    \includegraphics[width=1.0\columnwidth]{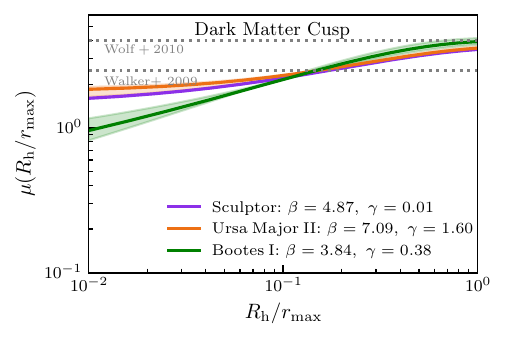} 
    \hspace{-4.2mm}
    \includegraphics[width=1.0\columnwidth]{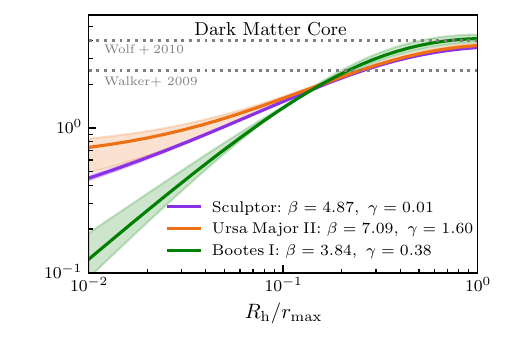} % Adjust width as needed % Adjust width as needed 
    \caption{Like Figure \ref{MFL fig}, but showing $\mu$ as a function of the ratio of projected stellar half-light radius to the characteristic size of the dark matter halo ($R_{\mathrm h} /r_{\mathrm{max}}$). Here, $\mu$ depends on $R_{\mathrm h} /r_{\mathrm{max}}$ as well as the shapes of the stellar and dark matter profiles. There is a significant variation in $\mu(R_{\mathrm h} /r_{\mathrm{max}})$ between different $2\beta\gamma$ profiles. The left (right) hand panel shows $\mu$ calculated under the assumption of a cuspy (cored) dark matter halo (see Equations \ref{cusp density} and \ref{core density}) . Solid lines show the best fit parameters, shading shows 1-sigma observational uncertainties on $\beta$ and $\gamma$. The two dotted grey lines show the mass estimators of \cite{walker2009universal} and \cite{wolf2010accurate} as labeled. Note that for this figure we treat stars as massless tracers.}
    \label{cusp/core}
\end{figure*}

Note that, in the previous subsection on mass follows light models, we consider how $\mu$ and therefore the enclosed mass estimate varies with enclosing radius $r_0$ and how $\mu$ differs between different stellar density profiles at the same $r_0$. In this section, both $\mu$ and the enclosed mass estimate are evaluated at the half-light radius, and we instead see how $\mu$ varies with different assumptions about the unknown properties of the dark matter halo and how $\mu$ differs between stellar density profiles within the same halo. Figure \ref{cusp/core} shows the distinct functional dependence of $\mu$ on $R_\mathrm{h}/r_\mathrm{max}$ for the three example galaxies introduced in Section \ref{sec data}) embedded in cupsy (left) and cored dark matter halos (right). There is significant uncertainty in $\mu$ due to ignorance of the segregation $R_\mathrm{h}/r_\mathrm{max}$ (as discussed in \citealt{errani2018systematics} for the case of Plummer profiles). We note that the shape of $\mu$ vs. $R_\mathrm{h}/r_\mathrm{max}$ clearly depends on the different shapes of the stellar density profile.

The best-fit $\beta$ and $\gamma$ of the Sculptor dwarf galaxy (purple curve in Figure \ref{cusp/core}, $\beta = 4.87\pm0.05,\,\gamma = 0.01\pm0.01$), which is well approximated by the Plummer model (used in most simple mass estimators, see, e.g.,\citealt{ walker2009universal, wolf2010accurate,errani2018systematics}) provides a useful contrast to the more extreme slopes of Ursa Major II (orange, $\beta = 7.09\pm1.62,\,\gamma = 1.60\pm0.09$) and Bootes I (green, $\beta = 3.84\pm0.22,\,\gamma = 0.38\pm0.26$). For a given stellar density profile, the systematic uncertainty in $\mu$ due to ignorance of $R_\mathrm{h}/r_\mathrm{max}$ is magnified in cored dark matter halos compared to cuspy ones (as discussed in \citealt{errani2018systematics}, see their Figure 1). We note that $\mu$ is especially sensitive to the outer slope $\beta$. For example, for a fixed range of $0.01 \leq R_\mathrm{h}/r_\mathrm{max} \leq1$, the Ursa Major II dwarf galaxy (orange curve in Figure \ref{cusp/core}), spans a smaller range of $\mu$ compared to the Plummer-like Sculptor. Its steep outer slope (large $\beta$) tells us that the stellar density drops off quickly outside the scale radius. The more steeply the stellar component is truncated in the outskirts, the less sensitive the mass estimate is to the relative extent of the dark matter halo. On the other hand, in systems for which a shallow outer slope in stellar density (small $\beta$) is observed, such as the Bootes I dwarf galaxy (green), the systematic uncertainty in $\mu$ is amplified, allowing for an uncertainty of over an order of magnitude on the enclosed mass. 

We find that $\mu$ peaks around $R_\mathrm{h}/r_\mathrm{max}\sim1$ and, for $R_\mathrm{h}/r_\mathrm{max}>>1$,
asymptotically approaches a value of 1.88 (2.50, 2.99) for the case of Ursa Major II
(Bootes I, Sculptor). Note that this asymptotic value is independent of the inner
halo shape (cusp/core), but is sensitive to the details of the stellar
density profile.

Significant uncertainty on $\mu$ can remain even if $R_\mathrm{h}/r_\mathrm{max}$ were known: at a fixed segregation, $\mu$ remains sensitive to the choice of stellar density profile and fitted parameters. This is especially important at small $R_{\rm h}/r_{\rm max}.$ Figure \ref{cusp/core} highlights the importance of the stellar density profile on the dynamical mass estimate, especially in those systems well-fit by shallow outer density slopes and that may be embedded in cored dark matter halos. 

\subsection{$\mu$ and $R_{\text{h}}$ for the Milky Way dwarfs}
\label{mu and rh}
The mass given by Equation \ref{mass 2} depends critically on the quantities $\mu$ and $R_{\text{h}}$ and their associated uncertainties. Here we examine how the value of $\mu$ depends on the shapes of the stellar and dark matter density profiles and the segregation $R_{\text{h}}/r_{\text{max}}$ as described in Section \ref{stars + dm} and how $R_{\text{h}}$ depends on the form of the fitted stellar density profile. We find that allowing a flexible functional form for the stellar density profile significantly expands the range of possible $\mu$ values and hence the range of viable dynamical masses.
\subsubsection{Fitted values of $R_{\text{h}}$ for the Milky Way dwarfs}
\label{rh subsubsection}

\begin{figure*}[t] % 'h' means "here"
    \centering
    \includegraphics[width=1.0\textwidth]{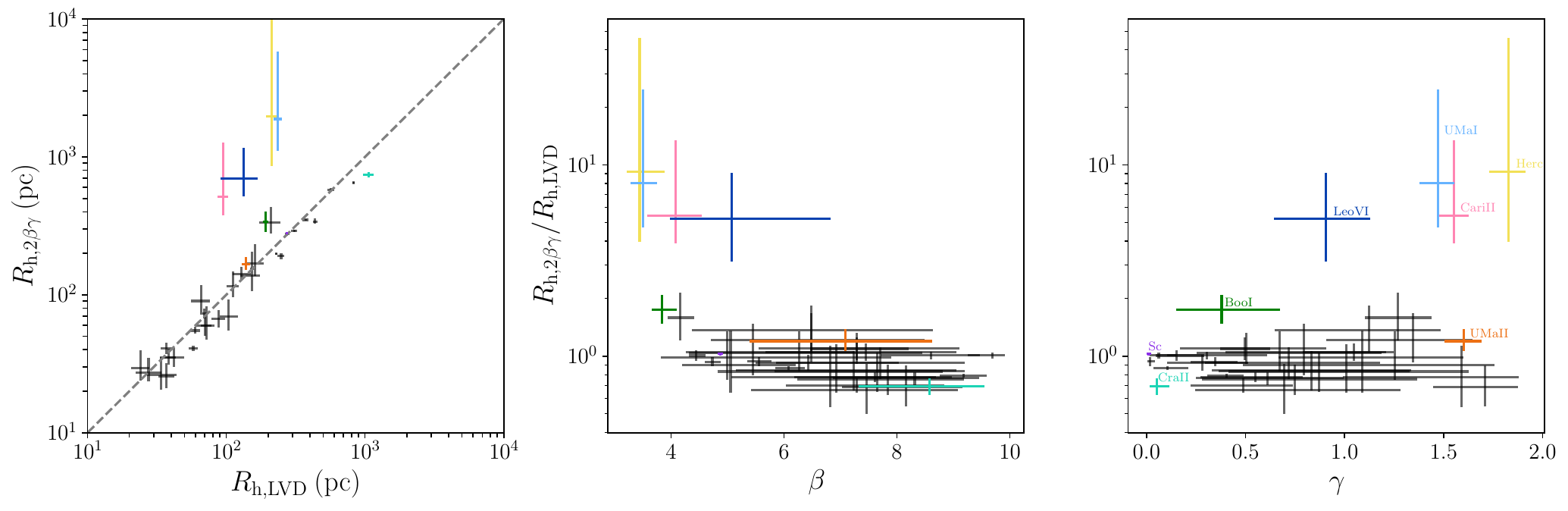} 
    \caption{The half-light radii obtained from the fits to the $2\beta\gamma$ model described in Section \ref{sec data} are in most cases consistent with published values in the Local Volume Database (LVD) of \cite{Pace2024arXiv241107424P}. The left-hand panel compares the half-light radii from the LVD against our fits, revealing four outliers:  Hercules (yellow), Ursa Major I (light blue), Carina II (pink), and Leo VI (dark blue) obtained by fitting $2\beta\gamma$ models with half-light radii $5-10$ times larger than previously published values from the LVD.  The center and right-hand panels show the ratio of the two half-light radii as a function of the fitted value of $\beta$ and $\gamma$, respectively. Sculptor, Bootes I, and Ursa Major II are plotted in the same colors as in Figures \ref{SB fig}-\ref{cusp/core}, and Crater II is shown in teal. The dashed line on the leftmost panel shows $R_{\mathrm h, 2\beta\gamma} = R_{\mathrm{h,LVD}}$.}
    \label{rh fig}
\end{figure*}

Figure \ref{rh fig} compares our $2\beta\gamma$ fits
of the stellar half-light radius $R_{\mathrm h, 2\beta\gamma}$ with the literature
half-light radii $R_{\mathrm{h,LVD}}$ as compiled in the LVD \citep{Pace2024arXiv241107424P}, and shows how this ratio depends on the fitted $\beta$ and $\gamma$. For most galaxies, our fitted half-light radii are consistent with those in the LVD, with 26 of the 36 galaxies differing by less than $25\%$. However, there are a couple of systems (Hercules and Ursa Major I) whose $2\beta\gamma$ half-light radii are almost an order of magnitude higher (factors of 9 and 8, respectively) than the half-light radii listed in the LVD. The other outliers are Carina II and Leo VI, which have $R_{\mathrm h, 2\beta\gamma} \sim 5 R_{\mathrm{h,LVD}}$. Due to the intrinsic correlation between $R_{\mathrm h}$ and $M_{\star}$ in the region of the posterior with $\beta \lesssim  5$, they also have larger total stellar mass in the $2\beta\gamma$ fits relative to the LVD. This results in approximately an order of magnitude lower average surface brightness, as this quantity goes as $M_\star/R_{\mathrm h}^2$.  We notice that all four of objects have very shallow 
$\beta$, cuspy $\gamma\sim2$ tending towards the upper bound of the prior in our $2\beta\gamma$ fits, and observed luminosities of $\lesssim 10^5 \, L_{\odot}$ \citep{Pace2024arXiv241107424P}. As explained in Paper II, when the stellar density favors a cuspy profile with $\gamma\sim 2$ and $\beta$ is relatively unconstrained, as is the case with these galaxies, then the prior volume at $\beta<5$ extends toward large $R_{\rm h}$ as the outer extent of the galaxy is poorly constrained. In the case of Hercules, the deviation between the $2\beta\gamma$ and LVD results is a consequence of our spherical fits: the system is manifestly flattened, which leads to shallow $\beta$ and inflated $R_{\rm h}$ when fit with the spherical models. However, this same effect does apply to the other three outliers discussed above. Due to the importance of $R_{\rm h}$ on the mass estimate, these effects further motivate consideration of the unique $\beta$ and $\gamma$ of each galaxy.

\subsubsection{Systematics in $\mu$ for the Milky Way dwarfs}
\label{mu subsubsection}

\begin{figure*}[t] % 'h' means "here"
    \centering
    \includegraphics[width=1.0\textwidth]{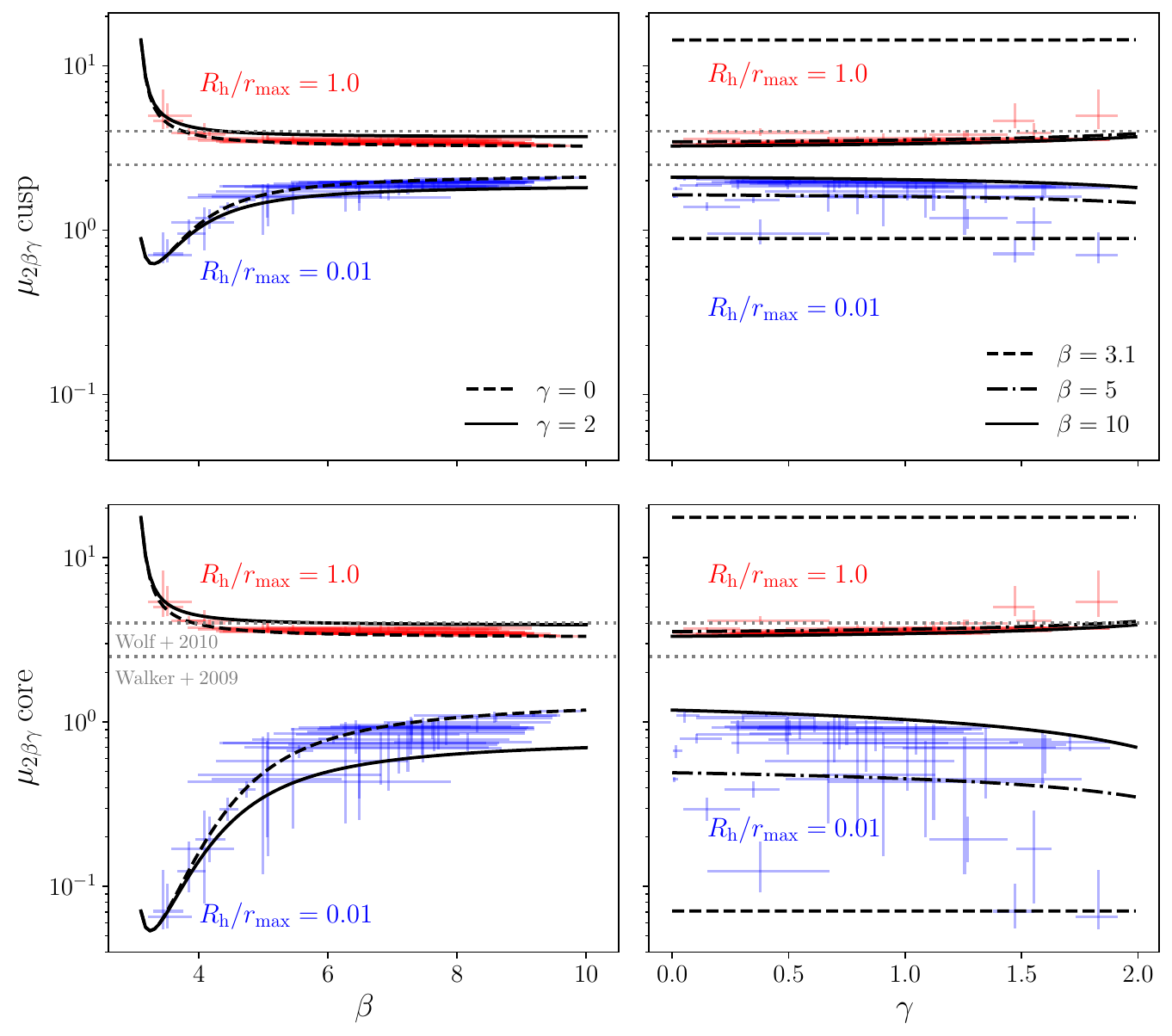} 
    \caption{The dependence of $\mu$ on the inner and outer slopes $\beta$ and $\gamma$ of the $2\beta\gamma$ stellar density profile, for cuspy (upper row) and cored (lower row) dark matter halos. We show in blue and red the values of $\mu$ from the best-fit $2\beta\gamma$ model for each dSph for a fixed segregation of $R_{\mathrm{h}}/r_{\mathrm{max}} = 0.01$ and $1.0$, respectively. Error bars show propagated observational uncertainties. Black curves are computed directly from Equation \ref{mu 1}, with different line styles corresponding to different fixed values of $\beta$ and $\gamma$ (left and right, respectively). The range of $\beta$ and $\gamma$ shown by the black curves is the same as the prior range on those parameters in the fits described in Section \ref{sec data}. The two dotted grey lines show the mass estimators of \cite{walker2009universal} and \cite{wolf2010accurate} as labeled.}
    \label{mu figure}
\end{figure*}

As shown in the previous section, the ignorance of segregation discussed is a source of uncertainty in dSph mass estimates that cannot be neglected. In the following, generalizing the conclusions of Figure \ref{cusp/core} to the Milky Way dSph population as a whole, we show that this systematic uncertainty is highly sensitive to the inner and outer slopes $\beta$ and $\gamma$ of the stellar density profile. In many cases, the uncertainty is magnified relative to the Plummer model, highlighting the importance of each galaxy's unique stellar density profile to the mass estimate. The observed properties of the Milky Way satellite galaxies demand a wide range of $\mu$: this is illustrated in Figure \ref{mu figure}, where we show the functional dependence of $\mu$ on $\beta$ and $\gamma$ for example segregations $R_{\mathrm{h}}/r_{\mathrm{max}}$ of $0.01$ and $1.0$. The top and bottom panels display the results for cuspy Hernquist (Equation \ref{cusp density}) and cored Dehnen dark matter halos (Equation \ref{core density}), respectively. We show in red and blue the values of $\mu$ calculated directly from Equation \ref{mu 1} for the best fit $\beta, \gamma$ of each dSph in our dataset assuming a segregation $R_{\mathrm{h}}/r_{\mathrm{max}}$ of $1.0$ and $0.01$, respectively. The error bars show the spread of $\mu$ due to observational uncertainty in the model parameters $\beta$ and $\gamma$, corresponding to the 16th to the 84th percentile of the underlying Monte-Carlo sampled distribution. To guide the eye, the black lines show the dependence of $\mu$ from Equation \ref{mu 1} at fixed segregations and fixed values of $\beta$ and $\gamma$. The dashed and solid lines show the lower and upper bounds, respectively, on the priors on either $\beta$ and $\gamma$ depending on the plot (see Section \ref{sec data}). The dashed line on the left is the Plummer value of $\gamma=0$, while on the right, the dash-dotted line represents the Plummer value of $\beta =5$. The grey dotted lines show the value of $\mu$ in the \cite{walker2009universal} and \cite{wolf2010accurate} mass estimators, as labeled. 

Figure \ref{mu figure} highlights three main lessons on how the stellar density profile, the dark matter halo, and their relative sizes affect the estimate of enclosed dynamical masses for the Milky Way dSph population. 

$\bullet$ $\mu$ is sensitive to the outer slope $\beta$, magnifying the previously studied uncertainties due to unknown segregation and the shape of the dark matter halo. At fixed segregation, $\mu$ depends more sensitively on $\beta$ than on $\gamma$. The sensitivity to $\beta$ is especially evident when the outer stellar density profile is shallower than a Plummer profile ($\beta < 5$), motivating observations of the outskirts of these systems to constrain their outer density slopes.

$\bullet$ Regardless of the unique stellar density profile, the range of $\mu$ is magnified when a galaxy occupies a cored dark matter halo rather than a cuspy one, as anticipated by Figure \ref{cusp/core}. Additionally, the sensitivity of $\mu$ on $\beta$ is amplified in cored halos compared to cuspy ones. This is important to consider, as there are few systems with well-constrained $\beta$.

$\bullet$ Further, we note that the more deeply a galaxy is embedded within its dark matter halo (smaller $R_{\mathrm h}/r_{\mathrm{max}}$), the more sensitive $\mu$ is to the outer stellar density slope $\beta$.

In summary, \textit{the uncertainty in the mass estimate depends systematically on the stellar density profile.} The observed diversity of these profiles, especially their outer slopes $\beta$, intensifies all known uncertainties due to unknown segregation and shape of the dark matter profile. The most widely used simple mass estimators (dotted grey lines) systematically undershoot the mass for systems with small values of segregation and shallow $\beta$, challenging the approximation of applying the same $\mu$ to all dSphs.

\subsection{Including the contribution of stellar mass to the potential}
\label{sec:not massless}
In addition to incorporating the unique shape of the stellar density profile, the virial theorem allows a formalism for including the stellar contribution to the  enclosed dynamical mass. This enables us to more accurately reflect the total mass of the system, especially in galaxies like Fornax and Leo I where the stellar contribution to the total mass is significant (for example, within the half-light radii their dynamical mass to light ratios are estimated to be ${\approx}10\,(M/L)_{\odot}$ and ${\approx}8\,(M/L)_{\odot}$, respectively) \citep{Pace2024arXiv241107424P}. First, we separate the total dynamical mass into its two distinct components, dark matter and baryons:

\begin{equation}
    M_{\mathrm{tot}}({<}r) = M_{\text{DM}}({<}r) + M_{\text{bar}}({<}r)
\label{total mass}
\end{equation}

where the enclosed baryonic mass is

\begin{equation}
    M_{\text{bar}}({<}r) = M_\star \int_0^r 4\pi s^2\nu_\star(s) ds,
\label{baryonic mass}
\end{equation}

$M_\star$ being the fitted total stellar mass and $\nu_\star$ the fitted $2\beta\gamma$ profile. In order to calculate $M_{\text{DM}}({<}R_{\text{h}} )$, we use Equation \ref{mass 1}

\begin{equation}
    M_{\text{DM}}({<}R_{\text{h}}) = \frac{\mu R_{\text{h}} \langle \sigma^2_{\text{los, DM}} \rangle}{\mathrm G},
\label{dm mass}
\end{equation}

where $\langle \sigma^2_{\text{los, DM}}\rangle$ is the contribution to the velocity dispersion driven by the underlying dark matter potential. To compute this value, we separate the observed velocity dispersion into its stellar and dark matter components. First, we calculate the dispersion sourced only by baryons using the virial theorem (see Equations \ref{virial} - \ref{eq:wlos}): 
\begin{equation}
\langle \sigma_{\text{los, bar}}^2 \rangle = W_{\text{los, bar}},
\end{equation}
where $W_{\text{los, bar}}$ is calculated by setting $M({<}r) = M_{\text{bar}}({<}r)$ in Equation \ref{eq:wlos}. We then subtract $\langle \sigma_{\text{los, bar}}^2 \rangle$ from the total $\langle \sigma_{\text{los}}^2 \rangle$ to obtain the excess dispersion, attributed to dark matter:

\begin{equation}
\langle \sigma_{\text{los, DM}}^2 \rangle = \langle \sigma_{\text{los}}^2 \rangle - \langle \sigma_{\text{los, bar}}^2 \rangle.
\end{equation}

This procedure allows us to estimate the total dynamical mass enclosed within the half-light radius, taking into account the contribution of stellar mass. We also avoid propagating the significant uncertainties in $\mu$ to the contribution from baryonic mass, which can be estimated directly (see Section \ref{sec data}). For the Milky Way dwarf Spheroidal galaxies Leo~II, Carina, and Sculptor, which have the largest contribution from stellar mass, we find that $\langle \sigma^2_{\mathrm{bar}}\rangle/\langle\sigma^2_{\mathrm{los}}\rangle \approx 2.8\%, 2.3\%$ and $2.1\%$, respectively; hence, the inclusion of stellar mass represents a percent-level correction to the total mass estimates for these objects.

\section{Application to scaling relations}
\label{sec scaling relations}

Comparison of observed and simulated scaling relations acts as an empirical test of competing theories of dark matter, cosmology, and modified gravity. Key examples include dynamical density vs. radius, the stellar mass-halo mass relation, the size-mass relation, and the mass-concentration relation \citep{2012AJ....144....4M, 2019ARA&A..57..375S}. These scaling relations can be used to probe the structure of dark matter subhalos \citep{salucci2000dark}, star formation history \citep{behroozi2013average}, baryonic feedback processes \citep{sales2022baryonic}, and environmental effects such as tidal perturbations \citep{watkins2023possible}. 

Due to the dependence on simple mass estimators inherent to the study of dwarf Spheroidal galaxies, these dynamical scaling relations depend on the shape of the stellar density profile and the assumptions used to relate the enclosed mass $M({<}r)$ to velocity dispersion $\langle \sigma^2_{\mathrm{los}} \rangle$ and 2D half-light radius $R_\mathrm{h}$ (see Equation \ref{mass 1}). In order to help the reader appreciate the extent of the uncertainties discussed in Section \ref{sec results}, we reexamine two scaling relations using the mass $M_{\mathrm{tot}}({<}R_{\rm h})$ given by Equation \ref{total mass}, tailored according to the $2\beta\gamma$ fit for each individual dSph: the total mass - stellar mass relation and the Radial Acceleration Relation. We choose these as examples and emphasize that our results have implications for \textit{any} dynamical scaling relations that rely on simple mass estimators. We see from Equation \ref{total mass} that $M_{\mathrm{tot}}({<}R_{\rm h})$ depends directly on the dimensionless factor $\mu$, and hence on the systematics discussed in Section~\ref{sec results}: for a fixed dark matter halo shape and a particular $2\beta\gamma$ stellar density profile, $\mu$ still depends on the segregation $R_\mathrm{h}/r_\mathrm{max}$. We remind the reader that this parameter is, a priori, unknown.

\subsection{Enclosed dynamical masses of the Milky Way dwarf spheroidals}
\label{enclosed mass relation}
\begin{figure*}[t] % 'h' means "here"
    \centering
    \includegraphics[width=1.0\columnwidth]{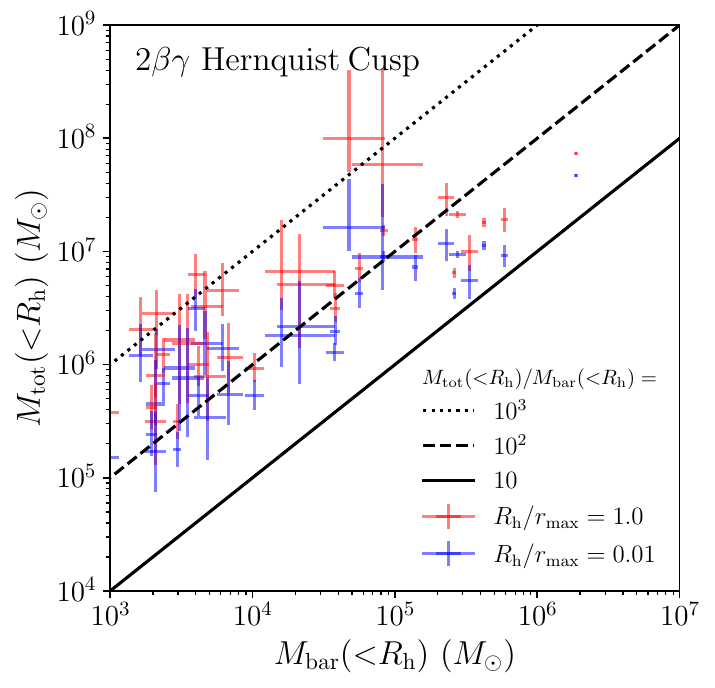} 
    \hspace{-3mm}
    \includegraphics[width=1.0\columnwidth]{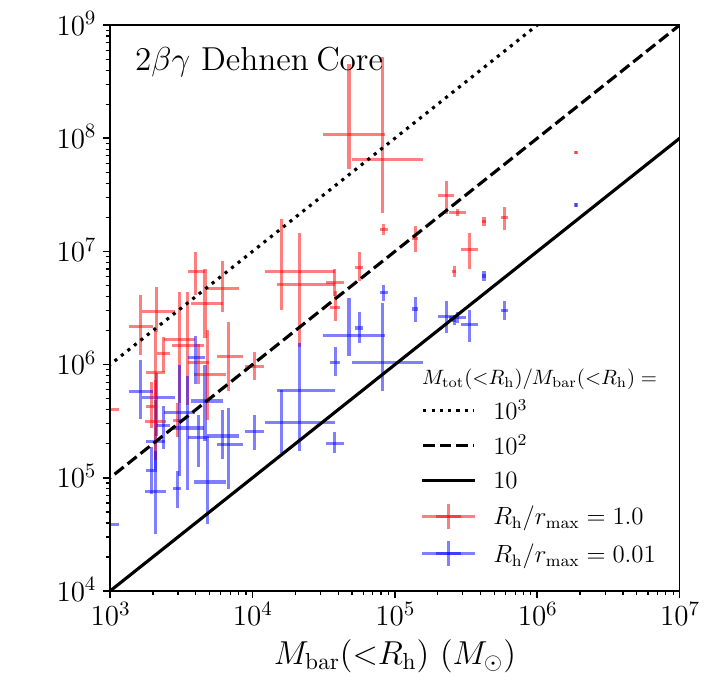} % Adjust width as needed % Adjust width as needed 
    \caption{Enclosed dynamical mass $M_{\mathrm{tot}}({<}R_{\rm h})$ of Milky Way dSphs computed using Equation \ref{total mass}, shown as a function of baryonic mass $M_{\mathrm{bar}}({<}R_{\rm h})$. Results for cuspy and cored dark matter halos are shown on the left and right, respectively. We show in blue and red the values of $M_{\mathrm{tot}}({<}R_{\rm h})$ for fixed segregations of $R_{\mathrm{h}}/r_{\mathrm{max}} = 0.01$ and $1.0$, respectively. The 1-sigma error bars depict observational uncertainty in the galaxy parameters, distance, and velocity dispersion. To guide the eye, we show lines of constant dynamical-to-stellar mass $M_{\mathrm{tot}}({<}R_{\rm h})/M_{\mathrm{bar}}({<}R_{\rm h})=$ $10$ (solid), $10^2$ (dashed), and $10^3$ (dotted). Note that the ratio $M_{\mathrm{tot}}({<}R_{\rm h})/M_{\mathrm{bar}}({<}R_{\rm h})$ depends sensitively on the assumed value of $R_{\rm h}/r_{\mathrm {max}}$.}
    \label{enclosed mass relation}
\end{figure*}

 Figure \ref{enclosed mass relation} shows the relationship between the enclosed total dynamical mass $M_{\mathrm{tot}}({<}R_{\rm h})$ and stellar mass $M_{\mathrm{bar}} ({<}R_\mathrm h)$ of Milky Way dwarf spheroidal galaxies. $M_{\mathrm{tot}}({<}R_{\rm h})$ is calculated using Equation \ref{total mass}, which depends on $\mu$ through $M_{\mathrm{DM}}({<}R_{\rm h})$ (see Equation \ref{dm mass}). For the stellar mass, we show $M_{\mathrm{bar}} ({<}R_\mathrm h)$ (Equation \ref{baryonic mass}) as fitted to the dSphs using $2\beta\gamma$ profiles. Velocity dispersions are taken from the LVD. Error bars show propagated uncertainties in the velocity dispersion, distance, half-light radius, and stellar mass, spanning from the 16th to the 84th percentile of the underlying Monte-Carlo sampled distribution. To address our ignorance of $R_\mathrm{h}/r_\mathrm{max}$, we plot $M_{\mathrm{tot}}({<}R_{\rm h})$ for fixed segregations of $R_{\mathrm{h}}/r_{\mathrm{max}} = 0.01$ and $1.0$ in blue and red, respectively. 

 We note that the ratio $M_{\mathrm{tot}}({<}R_{\rm h})/M_{\mathrm{bar}}({<}R_{\rm h})$ can vary by up an order of magnitude depending on the assumed segregation $R_{\mathrm{h}}/r_{\mathrm{max}}$. The largest variation between $R_{\mathrm{h}}/r_{\mathrm{max}} = 0.01$ and $1.0$ in both panels is Hercules. Assuming a cuspy (cored) dark matter halo, $M_{\mathrm{tot}}({<}R_{\rm h})/M_{\mathrm{bar}}({<}R_{\rm h})$ varies by a factor of 7 (72) for this system. We also note the next highest difference in $M_{\mathrm{tot}}({<}R_{\rm h})/M_{\mathrm{bar}}({<}R_{\rm h})$ between the two segregations is Bootes I with a factor of 4 (26) in the cusp (core) panel. This underscores a significant uncertainty in the properties and distribution of dark matter in the dwarf galaxy regime that is not captured by simple mass estimators. For a given panel of Figure \ref{enclosed mass relation}, depending on whether we take one or the other value of segregation, we would expect a drastically different behavior of the dark matter subhalo mass function in the low mass regime.

\subsection{A Re-examination of the Radial Acceleration Relation}
\begin{figure}[t] % 'h' means "here"
    \centering
    \includegraphics[width=1.0\columnwidth]{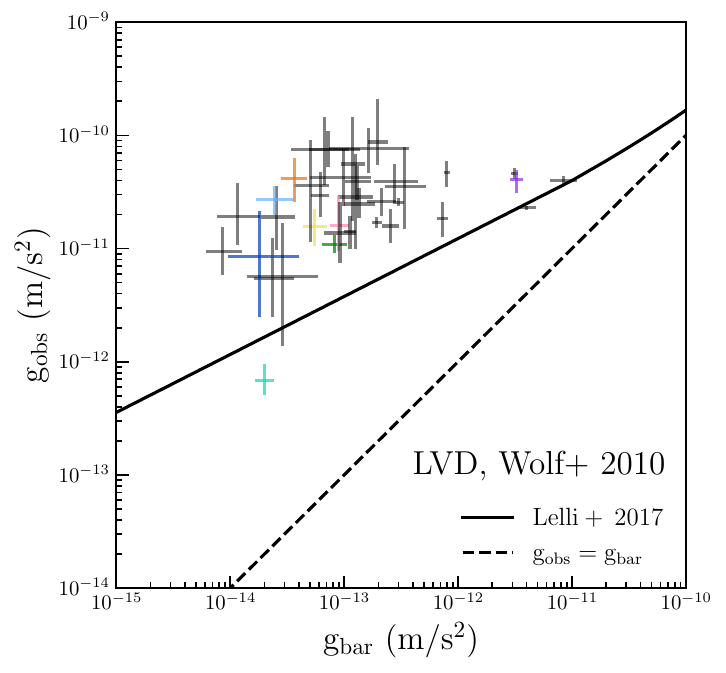} 
 % Adjust width as needed % Adjust width as needed 
    \caption{To ease comparison with Figure 10 of \cite{lelli2017one} in light of newly discovered systems and survey data, we here construct a RAR of Milky Way dSphs using a constant value of $\mu$. The total gravitational acceleration $g_\mathrm{obs}$ is computed using Equation \ref{mass 1} with the simple mass estimator $\lambda = 1$, $\mu = 4$ of \cite{wolf2010accurate}. The baryonic gravitational acceleration $g_\mathrm{bar}$ is derived from parameters of the stellar distribution from the Local Volume Database \citep{Pace2024arXiv241107424P}. The 1-sigma error bars depict observational uncertainty in the galaxy parameters, distance, and velocity dispersion. The dotted line shows $g_\mathrm{obs} = g_\mathrm{bar}$ and the solid black line is the fit to the Radial Acceleration Relation of late-type galaxies from \cite{lelli2017one}. Colors are the same as in Figure \ref{rh fig}. See Figure \ref{cusp core RAR} for a revised RAR that makes use of the virial theorem formalism and flexible $2\beta\gamma$ fits to showcase the results of Section \ref{sec results}}
    \label{LVD RAR}
\end{figure}

\begin{figure*}[t] % 'h' means "here"
    \centering
    \includegraphics[width=1.0\columnwidth]{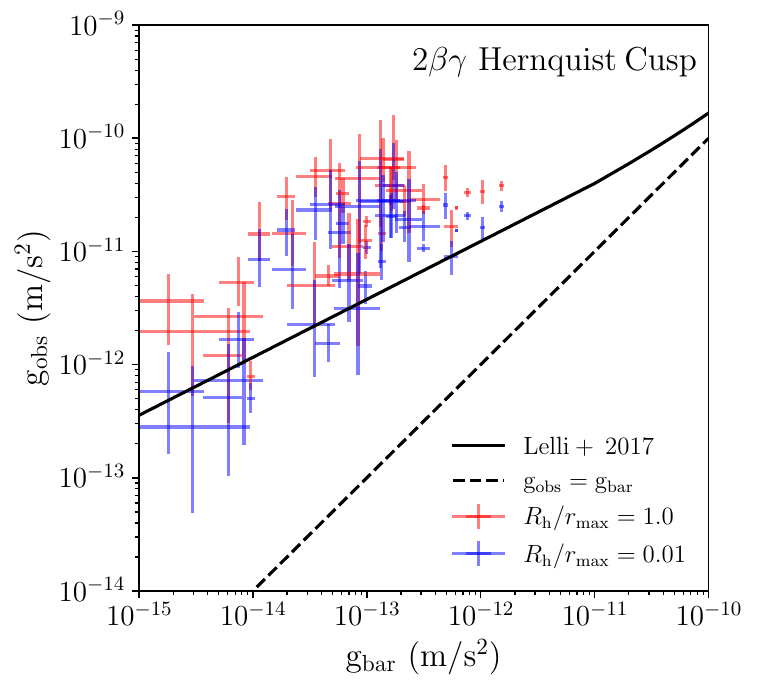} 
    \hspace{-3mm}
    \includegraphics[width=1.0\columnwidth]{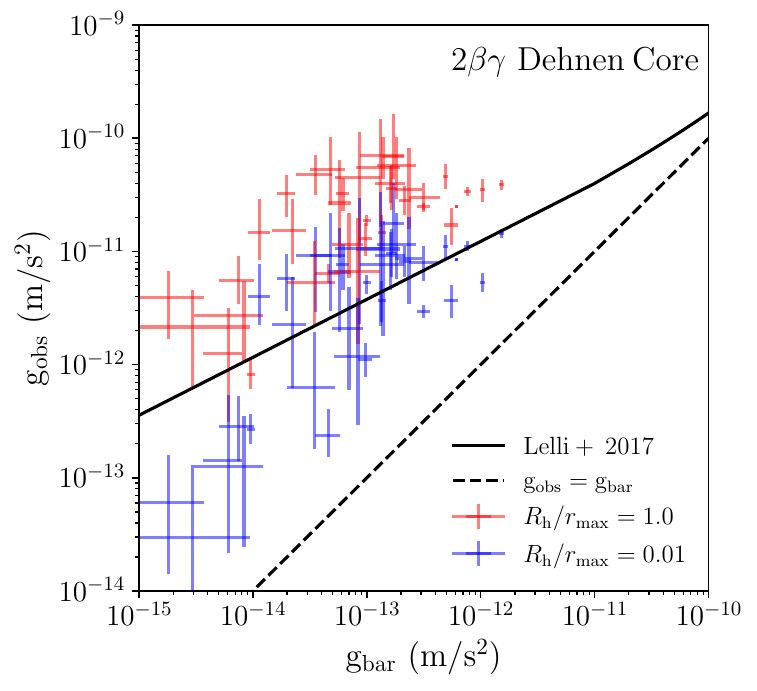} % Adjust width as needed % Adjust width as needed 
    \caption{Revised Radial Acceleration Relation of Milky Way dwarf spheroidal galaxies using flexible $2\beta\gamma$ fits and tailored dynamical mass from the virial theorem. The total gravitational acceleration $g_\mathrm{obs}$ is computed using Equation \ref{gobs} with $M({<}R_{\mathrm{h}})$ from Equation \ref{total mass} for cuspy and cored dark matter halos on the left and right, respectively. We show in blue and red the values of $g_{\mathrm{obs}}$ for fixed segregations of $R_{\mathrm{h}}/r_{\mathrm{max}} = 0.01$ and $1.0$, respectively. The baryonic gravitational acceleration $g_\mathrm{bar}$ is derived from  $2\beta\gamma$ surface brightness fits and is the same in both panels. The 1-sigma error bars depict observational uncertainty in the galaxy parameters, distance, and velocity dispersion. The dotted line shows $g_\mathrm{obs} = g_\mathrm{bar}$ and the solid black line is the fit to the Radial Acceleration Relation of late-type galaxies from \cite{lelli2017one}. }
    \label{cusp core RAR}
\end{figure*}

Next, we apply our tailored mass estimator to the Radial Acceleration Relation for Milky Way dwarf Spheroidals. The Radial Acceleration Relation (RAR) is a local scaling relation that connects the dynamical and baryonic accelerations as functions of radius, exhibiting a tight relation in late-type galaxies \citep{lelli2017one, mcgaugh2016radial}. \cite{lelli2017one} calculate the total gravitational acceleration $g_{\text{obs}}$ of late-type galaxies from the observed rotation curves, and the baryonic gravitational acceleration $g_{\text{bar}}$ from the observed radial density profiles of stars and gas. They find a tight correlation in the data for late-type galaxies, which is well fit by the following equation

\begin{equation}
    g_{\text{obs}} = \frac{g_{\text{bar}}}{1-e^{-\sqrt{g_{\text{bar}}/g_{\dagger}}}}
    \label{RAR eq}
\end{equation}

where $g_{\dagger} = [1.20\,\pm \, 0.02\, (\mathrm{stat})\pm 0.24 \, (\mathrm{sys})]\times 10^{-10}\, \mathrm m/ \mathrm s ^{2}$. They report scatter comparable to the observational uncertainties and residuals that are symmetric around zero, indicating no major systematics. \cite{lelli2017one} find early-type galaxies to be well described by Equation \ref{RAR eq} as well.

This tight correlation between the observed dynamical acceleration and the baryonic acceleration at the same radius in isolated objects is a basic tenet of Modified Newtonian Dynamics that follows from its fundamental equation \citep{milgrom1983modification}. Non-isolated objects such as satellite galaxies are subject to a varying External Field Effect and tidal effects, which may cause them to deviate from the RAR \citep{milgrom1983modification, 2000ApJ...541..556B, 2010ApJ...722..248M}. 

While the lowest baryonic acceleration probed by late-type galaxies in \cite{lelli2017one} is $g_{\text{bar}} \approx 10^{-12} \, \mathrm m/ \mathrm s ^2$, the dwarf galaxies probe accelerations as low as $10^{-14} \, \mathrm m/ \mathrm s ^2$ (see their Figure 10). The classical dwarfs are consistent with the RAR described by Equation \ref{RAR eq} to within their observational errors, while `ultrafaint' dwarfs display substantial scatter and near constant $g_{\text{obs}}$ over a wide range of $g_{\text{bar}}$. \cite{lelli2017one} find that this behavior persists even after applying cuts to remove low-quality data, although the scatter is reduced.

In light of new observational data, revised fits (see Section \ref{sec data} and Paper II), and the results of Section \ref{sec results}, we apply the virial theorem formalism to re-examine the RAR and show that both the dynamical acceleration $g_{\text{obs}}$ and the baryonic acceleration $g_{\text{bar}}$ are subject to systematic uncertainties.

We evaluate $g_\mathrm{obs}$ and $g_\mathrm{bar}$ at the half-light radius. The Newtonian acceleration $g_\mathrm{bar}$ expected from baryons alone is computed as
\begin{equation}
    g_\mathrm{bar} = \frac{G M_\mathrm{bar}({<}R_\mathrm{h})}{R_\mathrm{h}^2},
\label{gbar}
\end{equation}
where $M_\mathrm{bar}({<}R_\mathrm{h})$ is the stellar mass enclosed within the half-light radius. We compute $g_{\text{obs}}$ as follows
\begin{equation}
    g_{\text{obs}} = \frac{G M({<}R_{\text{h}})}{R_{\text{h}}^2},
\label{gobs}
\end{equation}
where $M({<}R_{\text{h}})$ is the enclosed dynamical mass, which will take different forms depending on the section.

\subsubsection{Updated RAR using LVD data and a simple mass estimator}
To provide an update to Figure 10 of \cite{lelli2017one}, before applying the formalism of Section \ref{sec results} to the RAR in Section~\ref{sec:VirialRAR}, we construct a version of the RAR which uses LVD data in combination with a \textit{simple mass estimator} for the calculation of $M({<}R_{\mathrm h})$ (\citealt{wolf2010accurate}: $\lambda = 1,\:\mu = 4$ in Equation \ref{mass 2}) in Equation \ref{gobs}. 
$R_{\mathrm h}$, stellar mass, and velocity dispersion are taken from the LVD to calculate $M_{\mathrm{bar}}({{<}R_{\mathrm h}})$ in Equation \ref{gbar}.
Specifically, $M_\mathrm{bar}({<}R_\mathrm{h})$ is calculated using the V-band absolute magnitude and distance taken from the LVD, assuming a mass-to-light ratio of $2\,M_{\odot}/L_{\odot}$, allowing a direct comparison with \cite{lelli2017one}. Error bars in Figure \ref{LVD RAR} show propagated uncertainties in the velocity dispersion, distance, half-light radius, and stellar mass, spanning from the 16th to the 84th percentile of the underlying Monte-Carlo sampled distribution. We include around 10 more Milky Way systems than \cite{lelli2017one}, some of which extend to lower $g_\mathrm{bar}$ than their sample, and find that the overall shape is similar whereas the scatter is somewhat reduced.

\subsubsection{Updated RAR using $2\beta\gamma$ fits and the virial theorem}
\label{sec:VirialRAR}
We will now construct a RAR that makes use of the \textit{tailored mass} given by Equation \ref{total mass}, which is shown Figure \ref{cusp core RAR}. For the computation of $g_\mathrm{bar}$, we use $R_\mathrm{h}$ and $M_\mathrm{bar}({<}R_\mathrm{h})$ (Equation \ref{baryonic mass}) as fitted to Milky Way dwarfs using $2\beta\gamma$ profiles (see Section \ref{sec data}). Velocity dispersions are taken from the LVD. For the computation of $g_\mathrm{obs}$, we insert $M_{\mathrm{tot}}({<}R_{\rm h})$ (Equation \ref{total mass}) into Equation \ref{gobs} to explicitly include the contribution of stellar mass to the potential. Under this operation, Equation \ref{gobs} becomes

\begin{equation}
    g_{\mathrm{obs}} = g_{\mathrm{bar}} + \frac{\mu\,\langle \sigma_{\text{los, DM}}^2 \rangle }{R_{\mathrm h}}.
    \label{gobs2}
\end{equation}
We see from the above equation that $g_{\text{obs}}$ depends directly on the dimensionless factor $\mu$. To address our ignorance of $R_\mathrm{h}/r_\mathrm{max}$, in Figure \ref{cusp core RAR} we plot $g_{\mathrm{obs}}$ for fixed segregations of $R_{\mathrm{h}}/r_{\mathrm{max}} = 0.01$ and $1.0$ in blue and red, respectively. The error bars are computed the same way as in Figure \ref{LVD RAR}. A comparison between Figures \ref{LVD RAR} and \ref{cusp core RAR} shows that the flexibility allowed by the $2\beta\gamma$ stellar density profile in combination with the tailored mass of Equation \ref{total mass} alters the shape and scatter of the RAR. In comparison, the application of a simple mass estimator in Figure \ref{LVD RAR} underestimates these systematic uncertainties. We will discuss the observed systematics on $g_{\mathrm{bar}}$ and $g_{\mathrm{obs}}$ separately in Sections~\ref{Sec:NewGBAR} and \ref{Sec:NewGOBS}.

\subsubsection{New systematics in $g_{\mathrm{bar}}$}
\label{Sec:NewGBAR}
As shown in Section \ref{sec data}, Milky Way dwarf spheroidals display a wide range of observed density profiles. A comparison between Figures \ref{LVD RAR} and \ref{cusp core RAR} shows that the flexibility allowed by $2\beta\gamma$ fits translates into a wider range of $g_{\mathrm{bar}}$. A consequence of this is the emergence of a new regime of ultra-low $g_{\mathrm{bar}} < 10^{-14} \: \rm{m}/\rm{s}^2$ in Figure \ref{cusp core RAR}, consisting of the 4 outlying systems discussed in Section \ref{rh subsubsection} in addition to Crater II. 

For the outlying systems (Carina II, Hercules, Ursa Major I, and Leo VI), the flexible $2\beta\gamma$ fits suggest average surface brightnesses $M_\star/R^2_{\mathrm h}$ a factor of $5-10$ lower than listed in the LVD. As $g_{\mathrm{bar}} \propto M_\star/R^2_{\mathrm h}$, these galaxies have up to order of magnitude lower $g_{\mathrm{bar}}$. Crater II has roughly opposite properties to the systems described above: it is best fit by a cored $2\beta\gamma$ profile and its half-light radius and stellar mass both increase relative to the LVD values, but in such a way that $g_{\mathrm{bar}}$ decreases by a factor of ${\sim2}$.  These results justify taking seriously the systematic uncertainty in $R_{\mathrm{h}}$ and $M_{\mathrm{bar}}({<}R_{\mathrm{h}})$ between different fitting profiles . 

\subsubsection{New systematics in $g_{\mathrm{obs}}$}
\label{Sec:NewGOBS}
As shown in Section \ref{mu subsubsection}, the flexibility of the $2\beta\gamma$ density profile exacerbates the uncertainties in $\mu$ for each system due to the unknown segregation $R_{\rm h}/r_{\mathrm{max}}$. A comparison between Figures \ref{LVD RAR} and \ref{cusp core RAR} shows that taking advantage of this flexibility to describe the diversity of observed density profiles reveals a significant systematic uncertainty in $g_{\mathrm{obs}}$. 

For each galaxy, there is a visible offset in the inferred $g_{\mathrm{obs}}$ depending on the assumed segregation $R_{\mathrm h}/r_{\mathrm{max}}$. In the left panel, showing the results for cuspy dark matter halos, this difference ranges from a factor of 2 (Leo I) to 7 (Hercules), depending on the galaxy. In the right panel, showing the results for cored dark matter halos, this factor ranges from 3 (Leo I) to 72 (Hercules). This ambiguity in $g_{\mathrm{obs}}$ leads to uncertainty in the overall shape and scatter of the RAR.  For a given panel of Figure \ref{cusp core RAR}, depending on whether we artificially fix the segregation to take the red or blue relation or some combination thereof, the data may either dramatically overshoot, undershoot, or closely follow Equation \ref{RAR eq}. What appeared as a flattening of $g_{\mathrm{obs}}$ in \cite{lelli2017one} may be an artifact of overly restrictive assumptions about the stellar and dark matter distributions; when the baryonic and dynamical accelerations are tailored to the individual stellar density profile, Milky Way dwarf galaxies appear consistent with the empirical RAR of Equation~\ref{RAR eq} even at low $g_{\mathrm{bar}}$.

\subsection{Discussion}
\label{sec discuss}
We emphasize that, while the observed flatness of the velocity dispersion profiles of the classical dSphs implies a constraint on the segregation for those systems \citep{walker2007velocity}, such constraints are absent for the less luminous and more prevalent fainter systems. Although there are significant variations in the results of Figures \ref{enclosed mass relation} and \ref{cusp core RAR} depending on the assumed segregation, we do not expect the simple vertical offset suggested by these figures to fully capture the systematic uncertainty. There is no clear prescription for the distribution of segregations throughout the Milky Way dSph population, and there is no clear reason to suggest that $R_{\rm h}/r_{\rm max}$ is fixed specifically at 0.01 or 1.0. However, various plausible values have been discussed in the literature. For example, cosmological simulations suggest that galaxies form deeply embedded within their hosts ($R_{\rm h}/r_{\rm max} \sim 0.01$ see, e.g.,  \citealt{2022MNRAS.510.3967R}, Figure 7 showing $M_{\star}\sim10^{7-9}M_{\odot}$ central galaxies; \citealt{2023MNRAS.520.1630K}, Figure 3 showing $M_{\star}>10^8M_{\odot}$ central galaxies; and \citealt{2025arXiv250203679S}, Figure 3 showing $M_{\star}\sim10^{9.75-11.25}M_{\odot}$ central and satellite galaxies). Once accreted onto a larger system like the Milky Way, $R_{\mathrm h}/r_{\mathrm{max}}$ can change under the influence of tides, with heavily stripped systems approaching a value of $\sim$ 1 \citep{errani2022structure}. Further, the fluctuating gravitational fields of subhalos inject energy into the stellar orbits, causing dSphs to expand \citep{2025arXiv250603904P}, altering the segregation. Finally, even segregations outside of $0.01 \leq R_\mathrm{h}/r_\mathrm{max} \leq1$ are allowed. This range is simply an estimate of the inherent uncertainty in segregation allowed by our current understanding. This ambiguity calls into question the overall structure of dynamical scaling relations for dwarf Spheroidal galaxies. In order to properly account for the effects of segregation on the enclosed dynamical mass, it would be necessary to model the distribution of segregation at the population level, which we plan to address in future contributions. Some scenarios (e.g., the formation of baryonic feedback induced cores \citealt{1996MNRAS.283L..72N}) allow for a mix of cored and cuspy dark matter halos in the dSph population, further increasing the scatter in these scaling relations. Framing a cuspy dark matter halo as one with a core size of zero, intermediate core sizes in between these extreme cases are also possible. 
We further note that our analysis was conduction under assumption of spherical symmetry and dynamical equilibrium. Both asphericity (see, e.g., \citealt{2018MNRAS.474.1398G}) and disequilibrium (see, e.g., \citealt{2025arXiv250418617T, 2025ApJ...992..162E}) are shown to affect dynamical mass estimates, further increasing the uncertainty of inferred masses.

Leveraging the $2\beta\gamma$ profile and virial mass estimator shows that \textit{the behavior of dwarf Spheroidals on dynamical scaling relations is subject to major systematics when a constant value of $\mu$ is assumed}. Specifically, there are knowable and striking systematics in $M_{\mathrm{bar}}({{<}R_{\rm h}})$ and $g_{\mathrm{bar}}({{<}R_{\rm h}})$, while $M_{\mathrm{tot}}({{<}R_{\rm h}})$ and  $g_{\mathrm{obs}}({{<}R_{\rm h}})$ systematics are less obvious and depend on the possible halo models considered. As it currently stands, the interpretation of dynamical scaling relations rests on these assumptions made in the mass estimate. We note that this uncertainty would be present in any dynamical variable and have implications for any such scaling relations.

\section{Conclusions}
\label{sec conclusion}

Dynamical mass estimators and the systematics present therein play a crucial role in the study of dark matter halos on the scale of dwarf Spheroidal galaxies. Previous work has shown systematic uncertainties in dSph mass estimators due to the unknown inner density slope and scale radius of the dark matter halo to be significant; however, the dynamical role of the stellar density profile is often overlooked. Assuming spherical symmetry and dynamical equilibrium within a given gravitational potential, the dSph's globally averaged velocity dispersion depends entirely on the shape of the stellar density profile. Tailoring the dynamical mass to the individual stellar density profile,
we show that both the inner and the outer slope of the
stellar density profile significantly affect the relationship between
globally-averaged velocity dispersion and enclosed mass. Because of this,
significant systematic uncertainties arise when assuming that the
density profile follows a rigid Plummer sphere, in other
commonly-used mass estimators. We conclude that these systematic uncertainties, implied by the observed diversity of dSph stellar density profiles, reach up to factors of order 10, significantly exacerbating all previously studied uncertainties in the enclosed dynamical mass. Including the modest contribution of stars to the potential, we apply our tailored dynamical masses to the enclosed dynamical mass - stellar mass relation and the Radial Acceleration Relation. We find significant uncertainties in their structure and scatter, highlighting the role of the individual stellar density profile in dynamical scaling relations. We advocate for using flexible $2\beta\gamma$ fits and modeling segregation at the population level to mitigate systematic error.

\begin{acknowledgements}
Large language models played no role in the writing of this manuscript.

This material is based upon work supported by the National Science Foundation Graduate Research Fellowship Program under Grant No. DGE2140739. Any opinions, findings, and conclusions or recommendations expressed in this material are those of the author(s) and do not necessarily reflect the views of the National Science Foundation.

\end{acknowledgements}

\bibliography{main}{}

\bibliographystyle{aasjournal}  % or apj, or aj

\end{document}